\documentclass[12pt]{article}

\usepackage{amssymb}
\usepackage{amsmath}

\begin{document}

\title{\bf Gravireggeons in Extra Dimensions and Interaction of
Ultra-high Energy Cosmic Neutrinos with Nucleons}

\author{A.V. Kisselev\thanks{E-mail: alexandre.kisselev@mail.ihep.ru} \
and V.A. Petrov\thanks{E-mail: vladimir.petrov@mail.ihep.ru} \\
\small Institute for High Energy Physics, 142281 Protvino, Russia}

\date{}

\maketitle

\thispagestyle{empty}

\begin{abstract}
We present the results on non-perturbative quantum gravity effects
related to extra dimensions which can be comparable, in some
cases, with the SM contributions, e.g. in lepton-lepton or
lepton-nucleon scattering. The case of cosmic neutrino
gravitational interaction with atmospheric nucleons is considered
in detail.
\end{abstract}
\vfill \eject

\section{Gravi-Reggeon effects in multidimensional scattering
amplidutes}
\label{sec:univ}

During last years there is a growing practical interest in models
with compact extra spatial dimensions. Their compactification
radius, $R_c$, varies from $1$ fm to $1$ mm, depending on a number
of extra dimensions $d=D-4$~\cite{Arkani-Hamed:98}. The models
predict massive Kaluza-Klein (KK) excitations of the graviton and
KK modes of the SM fields (provided the latter are allowed to
propagate in higher dimensions). If $D$-dimensional space-time has
a flat metric, the coupling of the massive graviton modes with the
SM particle is very weak and is defined by the Newton constant
$G_N = 1/\bar{M}_{Pl}^2$, where $\bar{M}_{Pl}$ is the reduced
Planck mass. Nevertheless, in the case when SM particles are
confined to a 4-dimensional flat ``brane'', summing up the KK
graviton excitations results in a $D$-dimensional gravitational
coupling $G_D \sim 1/M_D^{2 + d}$, with a fundamental Planck scale
$M_D$ of order $1$ TeV~\cite{Arkani-Hamed:98}.

Let us first consider the SM in \emph{$D$-dimensional flat
space-time}, $D>4$, \emph{without gravity}. Due to extra spatial
dimensions, an effective ``transverse interaction region'' becomes
larger than in four dimensions. One manifestation of this is a
modification of the Froissart-Martin upper bound in a \emph{flat}
space-time with arbitrary $D$ dimensions~\cite{Chaichian:92}:
\begin{equation}
\sigma_{tot}^D(s) \leqslant \text{const}(D) \,R_0^{D-2}(s),
\label{04}
\end{equation}
$\sqrt{s}$ being a collision energy. The ``transverse radius''  in
\eqref{04} is given by $R_0(s) = N(D) \ln s/\sqrt{t_0}$, where
$t_0$ denotes the nearest singularity in the $t$-channel, assumed
non-zero, while $N(D)$ is some integer depending on $D$. It is
interesting to see, what happens with scattering amplitudes when
we replace infinite extra dimensions by compact ones?

In Ref.~\cite{Petrov:01} the Froissart-Martin bound was
generalized for scattering in \emph{$D$-dimensional space-time
with compact extra dimensions}. For one extra dimension with the
compactification radius $R_c$, the upper bound is of the form:
\begin{equation}
\text{Im}\,T_D(s,0) \leqslant \text{const}(D)\,s
\,R_0^{D-2}(s)\,\Phi \left( \frac{R_0}{R_c},D \right),
\label{06}
\end{equation}
where $\text{Im}\,T(s,t)$  is the scattering amplitude, $t$ is a
momentum transfer (in $D$ dimensions) and $\Phi(R_0/R_c,D)$ is a
known function. At $R_c \ll R_0(s)$ the equality \eqref{06}
results in~\cite{Petrov:01}
\begin{equation}
\text{Im}\,T_D(s,0) \leqslant \text{const}(D)\,s \,R_0^{D-3}(s)\,
R_c,
\label{07}
\end{equation}
while in the opposite limit, $R_c \gg R_0(s)$, the inequality
\eqref{06} reproduces the upper bound \eqref{04}.

Now let us allow for the gravity to come into play. As was argued
in a number of papers~\cite{Amati:87}-\cite{Kabat:92}, in the
Minkowski space-time with $D$ dimensions ($D > 4$) the gravity
becomes strong in a transplanckian region ($s \gg M_D$), since an
effective gravitational coupling, $G_D s$, rises with energy.

In Refs~\cite{Amati:87} the eikonal representation for the
scattering amplitude of the gravitons in the string theory was
obtained:
\begin{equation}
A(s,t) = -2is \int d^{D-2} b  e^{i q b} \left [ e^{i \chi(s,b)} -
1 \right],
\label{08}
\end{equation}
where $\chi(s,b) \simeq \text{Im}\,\chi(s,b)$ is large at $b
\lesssim b_1 = 2 \sqrt{\alpha_g'} \ln s$ ($\alpha_g'$ is a string
tension). Thus, one get asymptotically
\begin{equation}
\sigma_{in}^D(s) \simeq \text{const}(D) \, b_1^{D-2}(s).
\label{10}
\end{equation}
Due to the absence of infrared divergences in the flat space-time
with more than four dimensions, the gravitational cross section
\eqref{10} appears to be finite and similar to the upper bound
\eqref{04}.

In what follows, we will first consider the scattering of two
particles in the model with \emph{one} compact extra dimensions
($D=5$) in the transplanckian kinematical region:
\begin{equation}
\sqrt{s} \gg M_D, \qquad s \gg -t,
\label{14}
\end{equation}
$t = -q_{\bot}^2$ being four-dimensional momentum transfer. The
generalization to $D > 5$ is straightforward and it will be done
below. Thus, we start from the consideration of the scattering of
bulk particles in four spatial dimensions, one of which is
compactified with the large radius $R_c$.

In the eikonal approximation an elastic scattering amplitude in
the transplanckian kinematical region \eqref{14} is given by the
sum of reggeized gravitons in $t$-channel. So, we assume that both
massless graviton and its KK massive excitations lie on linear
Regge trajectories:
\begin{equation}
\alpha(t_D) = \alpha(0) + \alpha'_g \, t_D,
\label{16}
\end{equation}
where $t_D$ denotes $D$-dimensional momentum transfer. Since the
extra dimension is compact with the radius $R_c$, we come to
splitting of the Regge trajectory \eqref{16} into a leading vacuum
trajectory~\cite{Petrov:02}
\begin{equation}
\alpha_0(t) \equiv \alpha_{grav}(t) = 2 + \alpha_g' t
\label{17}
\end{equation}
and infinite sequence of secondary, ``KK-charged'', gravireggeons:
\begin{equation}
\alpha_n(t) = 2 - \frac{\alpha_g'}{R_c^2}\,n^2 + \alpha_g' t,
\quad n \geqslant 1.
\label{18}
\end{equation}
The string theory implies that the slope of the gravireggeon
trajectory is universal for all $s$, and $\alpha_g' = 1/M_s^2$,
where $M_s$ is a string scale.

If we assume that multidimensional theory at short distances is a
string theory, than the scale $M_D$ can be of order the
fundamental string scale $M_s = (\alpha')^{-1/2}$. For instance,
in the type I theory of open and closed strings one
has~\cite{Antoniadis:98}
\begin{equation}
M_s = \left( \frac{g_s^2}{4\pi} \right)^{2/(2+d)} M_D,
\label{19}
\end{equation}
where $g_s$ is a gauge coupling at the string scale. This relation
leads to $D$-dimensional Planck mass a bit higher than the string
scale (for $g_s^2/4\pi \simeq 0.1$).

Thus, instead of taking a ``bare'' graviton exchange, we calculate
a contribution from the Pomeron trajectory to which this KK
graviton mode belongs:
\begin{equation}
- \, G_N \, \frac{1 + \exp(-i \pi \alpha_n(t))}{\sin \pi
\alpha_n(t)} \, \alpha_g' \beta^2_n(t) \left( \frac{s}{s_0}
\right)^{ \alpha_n(t)}. \label{20}
\end{equation}
The Born amplitude is, therefore, of the form
\begin{equation}
A^B(s,t,n) = G_N (2\pi R_c) \Big[ i - \cot \frac{\pi}{2}
\alpha_n(t) \Big] \alpha_g' \, \beta^2_n(t) \left( \frac{s}{s_0}
\right)^{ \alpha_n(t)}.
\label{22}
\end{equation}

In order to get an idea of possible $t$-dependence of Regge
residues $\beta^2_n(t)$, we consider scattering of $D$-dimensional
gravitons. The corresponding amplitude  has been calculated in
Refs.~\cite{Amati:87}:
\begin{equation}
A^B_{string}(s,t_D) \sim \frac{G_D \, s^2}{|t_D|} \,
\frac{\Gamma(1 - \alpha_g' t_D/2)}{\Gamma(1 + \alpha_g' t_D/2)} \;
(\alpha_g' s)^{\alpha_g' t_D}.
\label{24}
\end{equation}
The expression \eqref{24} is valid in the region $\alpha_g' |t| <
1$ in which it can be recast in the form:
\begin{equation}
A^B_{string}(s,t_D) \sim \frac{G_D \, s^2}{|t_D|} \, e^{\gamma \,
\alpha_g' t_D} \, (\alpha_g' s)^{\alpha_g' t_D},
\label{26}
\end{equation}
where $\gamma \simeq 0.58$ is the Euler constant.

Thus, we have $A(s,t) \sim \exp (\alpha_g' c \, t)$, where $c$ is
of order of unity. Let us assume that Regge residues in \eqref{22}
have an analogous $t$-dependence:
\begin{equation}
\beta^2_n(t) = \beta^2(0) \, e^{\alpha_g' b_0 (t - n^2/R_c^2)}.
\label{28}
\end{equation}
Since the coupling of all KK states to the SM fields is universal
in ADD model, we expect that $\beta^2_n(t)$ depends on $n$ via
$t_D = t - n^2/R_c^2$. Accounting for the fact that the product
$\alpha_g' b_0$ appears only in a combination with $\alpha_g' \ln
(s/s_0)$, we can neglect it in forthcoming  calculations at large
$s$ and put $\beta^2_n(t) \simeq \beta^2(0)$. At $t \rightarrow
0$, $n=0$ expression~\eqref{20} should reproduce  singular term
$G_N s/|t|$ related with the massless graviton, that results in
the relation $2 \beta^2(0)/\pi s_0^2=1$.

The expression for 5-dimensional eikonal amplitude looks like ($k$
being the exchanged KK quantum number)
\begin{equation}
A(s,t,k) = 2iR_c s \int d^2 b \; e^{i q_{\bot} b + i k \phi}
\int\limits_{-\pi}^{\pi} d \phi \left[ 1 -  e^{i \chi(s,b,\phi)}
\right], \label{30}
\end{equation}
with the eikonal given by
\begin{equation}
\chi(s,b,\phi) = \frac{1}{4\pi s} \int\limits_0^{\infty} q_{\bot}
d q_{\bot} \, J_o(q_{\bot} b) \,\frac{1}{2\pi R_c} \sum_{n= -
\infty}^{\infty} e^{- i n \phi}  A^B(s, -q_{\bot}^2,n).
\label{32}
\end{equation}
The variable $\phi$ runs the region $-\pi \leqslant \phi \leqslant
\pi$. These inequalities imply that $-\infty \leqslant y \leqslant
\infty$ in the limit $R_c \rightarrow \infty$ (flat extra
dimension), $y = R_c \phi$ being the 5-th component of the impact
parameter.

One can easily obtain from \eqref{30} that at $k=0$ and $s <
4/R_c^2$ only modes with $n=0$ contribute and effectively
$\chi(s,b,\phi) = \chi_0(s,b)$, corresponding to $n=0$
contribution in the sum in Eq.~\eqref{32}. So, at low energy the
scattering amplitude does not feel extra dimensions (the factor
$R_c$ is trivial and is absent at proper normalization).

Let us consider first the imaginary part of the eikonal. From
equations \eqref{32}, \eqref{22} we obtain:
\begin{equation}
\text{Im}\,\chi(s,b,\phi) = G_N s \, \frac{\alpha_g'}{8 R_g^2 (s)}
\exp \Big[ -b^2/ 4 R_g^2 (s)\Big] \,\theta_3(\upsilon,q),
\label{34}
\end{equation}
where
\begin{equation}
R_g(s) = \sqrt{\alpha_g' (\ln (s/s_0) + b_0)}
\label{36}
\end{equation}
is a gravitational slope. The quantity $\theta_3$ in \eqref{34} is
one of Jacobi $\theta$-functions~\cite{Bateman}:
\begin{equation}
\theta_3(\upsilon) = \theta_3(\upsilon,q) = 1 + 2
\sum_{n=1}^{\infty} \cos (2\pi n \upsilon) \,  q^{n^2}.
\label{38}
\end{equation}
In our case, it depends on variables
\begin{eqnarray}
\upsilon &=& \frac{\phi}{2\pi},
\nonumber \\
q &=& \exp \Big[ - R_g^2(s)/R_c^2 \Big].
\label{40}
\end{eqnarray}
The function $\theta_3(\upsilon,q)$ is well-defined for all
(complex) $\upsilon$ and all values of $q$ such as $|q| < 1$. It
has a singularity at $q \rightarrow 1$ (see below).
The $\theta$-functions are often defined in terms of variable
$\tau$:
\begin{equation}
\theta(\upsilon) = \theta(\upsilon|\tau),
\label{42}
\end{equation}
where
\begin{equation}
q = e^{i\pi \tau}.
\label{44}
\end{equation}

Let us define the ratio:
\begin{equation}
a = \frac{R_c}{2 R_g(s)}
\label{46}
\end{equation}
(that is, $q = \exp(-1/4a^2)$). From the
equality~\cite{Arkani-Hamed:98}
\begin{equation}
R_c = 2 \cdot 10^{31/d - 17}  \left( \frac{1 \text{TeV}}{M_D}
\right)^{1 + 2/d} \, \text{cm}
\label{48}
\end{equation}
we see that the compactification scale $R_c^{-1}$ varies from
$10^{-3}$ eV for $d=2$ to $10$ MeV for $d = 6$. Since $R_c^{-1}
\ll (2 R_g(s))^{-1}$ even at ultra-high energies, we have $a \gg
1$ and, consequently, $(1 - q) \ll 1$.

The behavior of $\theta_3(\upsilon,q)$ at $q \rightarrow 1$ can be
derived by using unimodular transformation of $\theta_3$-function
(known also as Jacobi imaginary transformation)~\cite{Bateman}:
\begin{equation}
\theta_3 \left( \frac{\upsilon}{\tau} \Big| -\frac{1}{\tau}
\right) = (-i\tau)^{1/2} e^{i\pi \upsilon^2/\tau}
\theta_3(\upsilon|\tau).
\label{50}
\end{equation}
Here $(-i\tau)^{1/2}$ has a principal value which lies in the
right half-plane. In variable $q$, equality \eqref{50} looks like
\begin{equation}
\theta_3(\upsilon,q) = \left( -\frac{\pi}{\ln q} \right)^{1/2}
\sum_{n=-\infty}^{\infty} e^{(2\pi n - \phi)^2/4\ln q}.
\label{52}
\end{equation}
The series in the RHS of \eqref{52} converges very quickly at $q
\rightarrow 1$, contrary to original series \eqref{38}:
\begin{equation}
\theta_3(\upsilon,q) = 2a \sqrt{\pi} \Big\{ e^{-\phi^2 a^2} +
\sum_{n=1}^{\infty} \Big[ e^{-(2\pi n - \phi)^2 a^2} + e^{-(2\pi n
+ \phi)^2 a^2} \Big] \Big\}.
\label{54}
\end{equation}
Notice that $a^2 = -1/4\ln q$.

From all said above, we get
\begin{equation}
\text{Im}\,\chi(s,b,\phi) \simeq G_N s  \, \frac{\alpha_g' R_c
\pi^{1/2}}{8 R_g^3(s)}  \exp \Big[ - (b^2 + R_c^2 \phi^2)/4
R_g^2(s)\Big].
\label{56}
\end{equation}

The expression \eqref{34} is directly generalized for $d$ extra
dimensions ($d \geqslant 1$):
\begin{eqnarray}
&& \text{Im}\,\chi(s,b,\phi_1, \ldots \phi_d) = G_N s \,
\frac{\alpha_g'}{8R_g^2(s)}
\nonumber \\
&& \times \exp \Big[ -b^2/4 R_g^2(s) \Big] \,\prod_{i=1}^d
\theta_3(\upsilon_i,q),
\label{58}
\end{eqnarray}
where $\upsilon_i=\phi_i/2\pi$. Correspondingly, we obtain
\begin{eqnarray}
&& \text{Im}\,\chi(s,b,\phi_1, \ldots \phi_d) \simeq G_N s \,
\frac{\alpha_g' R_c^d \pi^{d/2}}{8 R_g^{2 + d}(s)}
\nonumber \\
&& \times \exp \Big[ - (b^2 + R_c^2 \phi_1^2 + \ldots + R_c^2
\phi_d^2 )/4 R_g^2(s)\Big].
\label{60}
\end{eqnarray}
We see from \eqref{60} that the imaginary part of the eikonal
decreases exponentially in variables $b$, $\phi_i$ outside the
region:
\begin{equation}
b^2 + (R_c \phi_1)^2  + \ldots (R_c \phi_d)^2 \lesssim R_0^2(s),
\label{62}
\end{equation}
where
\begin{equation}
R_0^2(s) \simeq 4 R_g^2(s) \ln (s/M_D^2)
\label{64}
\end{equation}
at high $s$.

Let $t_D = (t, - n_1^2/R_c^2, \ldots, - n_d^2/R_c^2)$ be a bulk
momentum transfer. Then we get the following expression for
multidimensional scattering amplitude:
\begin{eqnarray}
A_D(s,t,n_1,\ldots,n_d) &=& -2 i s \, R_c^d \int d^2 b \; e^{i
q_{\bot} b} \int\limits_{-\pi}^{\pi} d \phi_1 \cdots
\int\limits_{-\pi}^{\pi} d \phi_d
\nonumber \\
&\times& \prod_{i=1}^d e^{i n_i \phi_i} \; \left[ e^{i
\chi(s,b,\phi_1,\ldots,\phi_d)} - 1 \right].
\label{66}
\end{eqnarray}
Correspondingly, the inelastic cross section in the space-time
with $d$ compact dimension is given by
\begin{equation}
\sigma_{in}^D(s) = (2 \pi R_c)^d \int d^2 b
\,\int\limits_{-\pi}^{\pi} d \phi_1 \cdots
\int\limits_{-\pi}^{\pi} d \phi_d \;  \Big[ 1 - e^{- 2\text{Im}\,
\chi(s,b,\phi_1,\ldots,\phi_d)} \Big].
\label{68}
\end{equation}
As was already shown, the imaginary part of the eikonal is
negligibly small outside region \eqref{62}. That results in the
estimates:
\begin{equation}
\sigma_{in}^D(s) \simeq \text{const}(D) \times \left\{
\begin{array}{ll}
R_0^{2 + d}(s),     & R_c \gg R_0(s)
\\ \\
R_0^2(s) \, R_c^d,  & R_c \ll R_0(s)
\end{array}
\right.
\label{70}
\end{equation}
which remind Eqs.~\eqref{04} and \eqref{07} obtained previously.
As was mentioned above, $R_0(s) \ll R_c$ for any reasonable $s$.
So, the size of the compact extra dimensions is irrelevant to the
behavior of the inelastic cross section and $\sigma_{in}^D(s) \sim
(\alpha_g')^{D/2-1} (\ln s)^{D-2}$. Only at $s \rightarrow
\infty$, when the transverse interaction region $R_0(s)$ becomes
much larger than $R_c$, we get $\sigma_{in}^D(s) \sim \alpha_g'
R_c^{D-4} (\ln s)^2$.

\section{Scattering of the SM fields in the presence of compact
extra dimensions}
\label{sec:brane}

Now we consider the case when \emph{the colliding particles are
confined on the 4-dimensional brane}, while the exchange quanta
(KK gravitons) are allowed to propagate in the bulk. Thus, the
collisions of the SM particles take place in a two-dimensional
impact parameter space. In Refs.~\cite{Giudice:02,Kisselev:03} the
scattering of two SM particles was calculated in the eikonal
approximation by summing up only ``bare'' KK gravitons. The
massive graviton modes originated from the extra dimensions change
four-dimensional propagator by
\begin{equation}
\frac{1}{-t} \; \rightarrow \sum_{n_1^2 + \ldots n_d^2 \geqslant
0} \frac{1}{-t + \sum\limits_{i=1}^d \displaystyle
\frac{n_i^2}{\displaystyle R_c^2}}.
\label{71}
\end{equation}
Since a contribution from only non-reggeized KK excitations of the
graviton has been taken into account
in~\cite{Giudice:02,Kisselev:03}, the eikonal has no imaginary
parts in such an approach. As was shown in~\cite{Kisselev:03}, the
$D$-dimensional brane amplitude has a renormalized Born pole at
$t=0$ and an infinite phase. Notice, series \eqref{71} diverges
and needs renormalization at $d \geqslant 2$.

In Ref.~\cite{Petrov:02*} it was shown that an amplitude of $M
\rightarrow N$ transition observed \emph{in four dimensions},
$A_{MN}$, is related to a corresponding $D$-dimensional amplitude
$A_{MN}^D$ by the relation
\begin{equation}
A_{MN} = (2\pi R_c)^{d(1 - (M+N)/2)} \, A_{MN}^D
\label{72}
\end{equation}
(in our case, $M=N=2$). Amplitudes $A_{MN}$ have non-zero limit at
$R_c \rightarrow 0$, reproducing the usual 4-dimensional
pseudoeuclidean case.  Since the colliding particles are confined
on the brane, their momenta lie in four-dimensional space.
Therefore, the impact parameter belongs to the two-dimensional
space and we have to put $\phi_i = 0$, $i=1, \ldots, d$, in
\eqref{60}. With taking account of this, the expression for
four-dimensional eikonal amplitude (in the presence of $d$ compact
extra dimensions) looks like
\begin{equation}
A(s,t) = 2i s \int d^2 b \; e^{i q_{\bot} b} \left[ 1 - e^{i
\chi(s,b)}\right],
\label{74}
\end{equation}
where $\chi(s,b) = \chi(s,b,\phi_1 = 0, \ldots, \phi_d = 0)$.
Taking into account \eqref{60}, we get the expression
\begin{eqnarray}
\text{Im}\,\chi(s,b) &=& \frac{1}{\pi^{d/2 -1}}
\, \frac{s}{ M_D^2} \left( \frac{M_s}{2M_D} \right)^d \, \Big[\ln
\Big( \frac{s}{s_0} \Big) \Big]^{-(1+d/2)}
\nonumber \\
&\times& \exp [ -b^2 /4 R_g^2(s)],
\label{76}
\end{eqnarray}
where the relation $M_{Pl}^2 = (2\pi R_c)^d \, M_D^{2+d}$ is used
~\cite{Arkani-Hamed:98}. The detailed analysis of the real part of
the eikonal will be given elsewhere. Here we only note that,
contrary to \eqref{76}, the real part of the eikonal (with the
massless graviton term subtracted) decreases as a power of the
impact parameter at large $b$.

The important features of the gravitational contribution to cross
sections are its independence of types of colliding particles and
a strong dependence on the collision energy . So, one can expect
that at superplanckian energies gravity exchanges will dominate
the SM electroweak interactions. That is why we now focus on
leptonic and semileptonic collisions.

Let us first consider $e^+e^-$ annihilation. Unfortunately, future
linear colliders will provide us with the c.m.s. energies
$\sqrt{s}$ around $M_D$ (0.5 $\div$ 2 TeV). In order to estimate
$\sigma_{in}^{e^+e^-}$ numerically, we need to fix Regge free
parameter $s_0$ in \eqref{76}. Since $s_0$ is related rather with
a mass scale of exchange quanta than with mass scales of colliding
particles, we can treat the scattering amplitude of two
graviton \eqref{24} instead of SM particle collision, and deduce that%
\footnote{In hadronic physics, the phenomenological parameter $s_0
\approx 1/\alpha'(0)$, where $\alpha'(0) \simeq 1$ GeV$^{-2}$ is
the slope of hadronic Regge trajectories~\cite{Collins:Regge}.}
\begin{equation}
s_0 = (\alpha'_g)^{-1}.
\label{78}
\end{equation}
This relation is also motivated by the
duality~\cite{Veneziano:68}. The results of our calculations of
inelastic cross section $\sigma_{in}^{e^+e^-}$ at $\sqrt{s} =
1$~TeV based on formulae \eqref{76},  \eqref{78} are presented in
the second row of Table~\ref{tab:e+e-}.
\begin{table}[h!t]
\begin{center}
\caption{{\small Cross sections of the processes induced by
graviton exchanges in $t$-channel (second row) and $s$-channels
(third row) at $\sqrt{s} = 1 \text{TeV}$ for different numbers of
extra dimensions $d$ (in \emph{pbarn}).}}
\bigskip
  \begin{tabular}{||c||c|c|c|c|c||}
  \hline $d$ & 2 & 3 & 4 & 5 & 6
  \\ \hline
  $e^+e^- \rightarrow e^+e^- + X$ & 1.06 $\cdot 10^3$ & 1.10
  $\cdot 10^2$ & 1.78 $\cdot 10^1$ & 3.84 & 1.02
  \\ \hline
  $e^+e^- \rightarrow f \bar f$ & 9.3 & 3.7 & 2.0 & 1.3 &
  0.9
  \\
  \hline
  \end{tabular}
\label{tab:e+e-}
\end{center}
\end{table}

These cross sections are larger than the cross sections of the
processes induced by massive graviton exchanges in $s$-channel (at
least, for $d \leqslant 6$).%
\footnote{It is worth to note that, generally, QFTs for $d > 0$ are
not renormalizable. So, the following estimates are of
illustrative character.}
For definiteness, consider matrix element for fermion pair
production $e^+e^- \rightarrow f \bar{f}$:
\begin{equation}
\mathcal{M} = G_N  T_{\mu\nu}^e P^{\mu\nu\alpha\beta}
T_{\alpha\beta}^f \, \sum_{n_1^2 + \ldots n_d^2 \geqslant 0}
\frac{1}{s - \sum\limits_{i=1}^d \displaystyle
\frac{n_i^2}{\displaystyle R_c^2}}.
\label{79}
\end{equation}
Here $P^{\mu\nu\alpha\beta}$ is a tensor part of a graviton
propagator, while $T_{\mu\nu}^{e(f)}$ is the energy-momentum
tensor of field $e(f)$~\cite{Giudice:99,Han:99}. The sum in
\eqref{79} diverges for $d \geqslant 2$. It can be estimated if
one convert it into an integral and introduce an explicit
ultraviolet cut-off $M_s$. Then we get for $d>2$:
\begin{equation}
\sum_{n_1^2 + \ldots n_d^2 \geqslant 0} \frac{1}{s -
\sum\limits_{i=1}^d \displaystyle \frac{n_i^2}{\displaystyle
R_c^2}} \simeq \left\{
\begin{array}{cl}
\displaystyle  - \frac{2 R_c^d}{(d - 2)\Gamma(d/2) (4\pi)^{d/2}}
\, M_s^{d-2}, &
\sqrt{s} \ll M_s \\ \\
\displaystyle  \frac{R_c^d}{\Gamma(1 + d/2) (4\pi)^{d/2}} \,
\frac{M_s^d}{s}, &
\sqrt{s} \gg M_s \\
\end{array}
\right.
\label{80}
\end{equation}
(an asymptotics at $\sqrt{s} \ll M_s$ was first found in
\cite{Han:99}).

Taking into account that the sum in indices results in a factor
proportional to $s^2$, we obtain from \eqref{78}, \eqref{80}:
\begin{equation}
\mathcal{M} \sim \lambda \, s^2 \, \Big( \frac{M_s}{M_D}
\Big)^{2+d} \times \left\{
\begin{array}{cl}
\displaystyle \frac{1}{M_s^4}, & \sqrt{s} \ll M_s \\ \\
\displaystyle \frac{1}{s M_s^2}, &  \sqrt{s} \gg M_s \\
\end{array}
\right.
\label{82}
\end{equation}
where $\lambda = \text{O}(1)$ has opposite sign for small and
large $\sqrt{s}$. Two asymptotics \eqref{82} are well-matched at
$\sqrt{s} \simeq M_s$. Thus, at $\sqrt{s} \gtrsim M_s$ we arrive
at the expression
\begin{equation}
\sigma(e^+e^- \rightarrow f \bar{f}) \simeq \lambda^2 \,
\frac{N_c}{40 \pi} \Big( \frac{M_s}{M_D} \Big)^{2+d}
\frac{s}{M_s^4},
\label{84}
\end{equation}
where $N_c$ represents the number of colors of the final state.
The result of numerical calculations by using formula \eqref{84}
is presented in the third row of Table~\ref{tab:e+e-}.

To compare, hadronic SM background in $e^+e^-$ annihilation
($e^+e^- \rightarrow e^+e^- + \text{ hadrons}$), including the
effects due to the (anti)tagging of the electron and accounting
for all available data on $\gamma \gamma$ collisions, was
estimated to be~\cite{Godbole:03}
\begin{equation}
\sigma_{had}^{e^+e^-}(\sqrt{s} = 1 \text{TeV}) \simeq (2.7 - 4.0)
\cdot 10^4 \, \text{pb}.
\label{86}
\end{equation}
The SM processes with different final states ($\sum_{q \neq t} q
\bar{q}$,\, $W^+W^-$, \, $t \bar{t}$, \, $\tilde{\chi^+}
\tilde{\chi^-}$, \, $\tilde{\mu^+_R} \tilde{\mu^-_R}$, \, $Zh$,
\emph{etc.}) have cross sections which are less than 1~pb at
$\sqrt{s} = 1$ TeV (see, for instance, Fig.~1.3.1 from
Ref.~\cite{Aguilar-Saavedra:01}). The highest rate has the process
$e^+e^- \rightarrow \sum_{q \neq t} q \bar{q}$, its cross section
is about 0.7~pb.

Our goal is to find the process in which gravity forces can
dominate SM interactions. Such a process has to obey the following
requirements: (i) colliding energy is much larger than $M_D \simeq
1$ TeV, (ii) SM cross section does not rise rapidly in $s$. The
best candidate is the scattering of ultra-high energy (UHE)
neutrinos off the nucleons. These neutrinos is a part of
ultra-high energy cosmic rays (UHECR) with energy $E \gtrsim
10^{18}$ eV, which are dominated by extragalactic sources of
protons~\cite{Bhattacharjee:00}. It is a detection of UHE
neutrinos that can help us to discriminate between different
origin of UHECR. For instance, in cosmological (``bottom-up'')
scenarios, neutrino fluxes are almost equal to gamma ray fluxes.
In astrophysical (acceleration) approach, the neutrino flux is
only a fraction of the gamma ray flux and is modified due to a
propagation of cosmic rays before they reach the Earth.

The cosmic neutrinos with extremely high energies $E \gtrsim
10^{20}$ eV are also believed to explain so-called
Greisen-Zatsepin-Kuzmin (GZK) cut-off of UHECR
spectrum~\cite{Greisen:66} (see below). During UHECR propagation,
the protons scatter off cosmic microwave background (CMB):
\begin{equation}
p + \gamma_{_{CMB}} \rightarrow N + \pi.
\label{88}
\end{equation}
Taking into account that a typical CMB photon energies are
$10^{-3}$ eV, one can obtain that the nucleon interaction length
drops to about 6 Mpc at GZK bound of $E_{GZK} \simeq 5 \cdot
10^{19}$ eV~\cite{Greisen:66}.%
\footnote{Below  $E_{GZK}$, the dominant energy loss for the
proton is due to the process $p \, \gamma_{_{CMB}} \rightarrow p
\, e^+ e^-$, down to the threshold energy of $4.8 \cdot 10^{17}$
eV.}
The observation of UHECR at $E > E_{GZK}$ is a serious problem for
theories in which the origin of CR is based on acceleration of
charges particles in astrophysical objects. Due to the energy
losses (say, through process \eqref{88}), the UHECR particles
cannot originate at distances larger than 60 Mpc from the Earth.
On the other hand, all potential astrophysical sources of UHECR
events are far beyond this distance.

At the same time the process \eqref{88} is an origin of so-called
cosmogenic neutrinos due to a consequence decay of charged pion as
$\pi^{\pm} \rightarrow \mu^{\pm} \, \nu_{\mu}$, $ \mu^{\pm}
\rightarrow e^{\pm} \, \nu_e \, \nu_{\mu}$. The fraction of the
proton energy carried by the neutrino is $E_{\nu}/E_p \approx
0.05$ and independent of $E_p$. The cosmogenic neutrino flux was
first estimated in \cite{Stecker:79}, \cite{Hill:85}. More recent
estimates can be found in Refs.~\cite{Stecker:91}-\cite{Engel:01}.
The predicted fluxes depend on the evolution parameter $m$ and on
the value of the redshift $z$, and lye in the range: $E_{\nu}^2 \,
\Phi_{\nu} \simeq ( 0.5 \cdot 10^{-9} - 10^{-8}) \, \text{ GeV}
\text{ cm}^{-2} \text{ s}^{-1} \text{ sr}^{-1}$  at $E_{\nu} =
10^{20} \, \text{eV}$ ($\nu = \nu_{\mu}, \bar{\nu}_{\mu}, \nu_e$).%
\footnote{Some cosmic ray protons with energies above 10$^{20}$ eV
are converted into neutrons by pion photo-production. The neutrons
decay again into protons during their propagation producing
electronic anti-neutrinos. This mechanism is important at
$E_{\nu_e} \lesssim 10^{17}$ eV.}

There are, however, other possible origin of UHE neutrinos. It is
usually anticipated that $\Phi_{\nu_e} \approx
\Phi_{\bar{\nu}_{\mu}} \approx \Phi_{\nu_{\mu}}$. We present below
the total flux of muonic neutrinos and antineutrinos in a number
of models at $E_{\nu} = 10^{20}$ eV. In the active galactic nuclei
(AGN), the dominant mechanism for neutrino creation is the
accelerated proton energy loss due to $pp$ or $p \gamma$
interactions~\cite{Mannheim:95}. Note, AGN produce a large
fraction of the gamma rays in the Universe, and their spectra
agree with the prediction that gamma rays are produced by hadrons.
In the AGN approach it was obtained that $E_{\nu}^2 \, \Phi_{\nu}
\simeq 0.3 \cdot 10^{-8} \, \text{ GeV} \text{ cm}^{-2} \text{
s}^{-1} \text{ sr}^{-1}$~\cite{Mannheim:95}. In Z-busts scenario,
cosmic neutrinos with extremely high energies ($E_{\nu} > 4 \cdot
10^{21} \, (1eV/m_{\nu})$ eV) collide with relic
neutrinos~\cite{Yoshida:98}. If the masses of the background
neutrinos $m_{\nu}$ are of several eV, the cosmic neutrinos
initiate high energy particle cascades which can contribute 10\%
to the observed cosmic ray flux at energies above the GKZ cut-off
(one of the main processes is a resonant $\nu \nu$ collision via
$Z$-bozon). The neutrino flux is $E_{\nu}^2 \, \Phi_{\nu} \simeq
0.3 \cdot 10^{-6} \, \text{ GeV} \text{ cm}^{-2} \text{ s}^{-1}
\text{ sr}^{-1}$~\cite{Yoshida:98}. In the so-called topological
defect models~\cite{Sigl:99}, UHECR are produced via decays of
supermassive $X$-particles related to a grand unification theory.
The expected neutrino flux is about $E_{\nu}^2 \, \Phi_{\nu}
\simeq 0.5 \cdot 10^{-6} \, \text{ GeV} \text{ cm}^{-2} \text{
s}^{-1} \text{ sr}^{-1}$~\cite{Sigl:99}. In gamma-ray busts (GRB)
model~\cite{Waxman:97}, the neutrino flux is strongly suppressed
at $E_{\nu} > 10^{19}$ eV, since the protons are not expected to
be accelerated to energies much larger than $10^{20}$ eV.

It is worth to mention model-independent upper bounds on the
intensity of high energy neutrinos produced by photo-meson
interactions. If the size of cosmic ray source is not larger than
photo-meson free path, the upper limit is (for evolving sources)
$4.5 \cdot 10^{-8} \, \text{ GeV} \text{ cm}^{-2} \text{ s}^{-1}
\text{ sr}^{-1}$~\cite{Bahcall:01}. However, for optically thick
pion photoproduction sources, the upper limit is less restrictive:
$2.5 \cdot 10^{-6} \, \text{ GeV} \text{ cm}^{-2} \text{ s}^{-1}
\text{ sr}^{-1}$~\cite{Mannheim:98}. Note, the considerably higher
flux of cosmogenic neutrinos was obtained in
Ref.~\cite{Kalashev:02}. The cosmogenic flux is the most reliable,
as it relies only on two assumptions: (i) the observed extremely
high energy cosmic rays (EHECR) contain nucleons, (ii) EHECR are
primarily extragalactic in origin.

One possible way to resolve the GZK puzzle%
\footnote{Note, however, recent paper~\cite{Bahcall:03}, in which
it is argued that the data from Fly's Eye, HiRes and Yakutsk
cosmic ray experiments are consistent with the expected
suppression of cosmic ray spectrum above $5 \cdot 10^{19}$ eV. The
AGASA data show an excess in this region.}
is to assume that the primary UHECR particles are neutrinos which
deposit a part of their energy to proton fragments in $\nu
N$-interactions. Unfortunately, the SM neutrino-nucleon cross
sections are not large enough to resolve the problem. Indeed, at
$10^{16} \, \text{eV} \lesssim E_{\nu} \lesssim 10^{21} \,
\text{eV}$ the conventional contributions from charged and neutral
current $\nu N$-scattering can be parameterized
by~\cite{Gandhi:96}
\begin{eqnarray}
&& \sigma_{\nu N}^{cc} \simeq 4.429 \cdot 10^3 \left(
\frac{E_{\nu}}{10^8 \, \text{GeV}} \right)^{0.363} \, \text{pb},
\nonumber \\
&& \sigma_{\nu N}^{nc} \simeq 1.844 \cdot 10^3 \left(
\frac{E_{\nu}}{10^8 \, \text{GeV}} \right)^{0.363} \, \text{pb}.
\label{90}
\end{eqnarray}
The total SM cross section for $\bar \nu N$-scattering has
practically the same magnitude and energy dependence at energies
under consideration ~\cite{Gandhi:96}. Putting, $E_{\nu} = 10^{21}
\, \text{eV}$ in \eqref{90}, we get an estimate $\sigma^{\nu
N}_{SM} \simeq 3.55 \cdot 10^5 \text{ pb}$. Such a value of
neutrino-nucleon cross section is two small to be relevant to the
GZK problem.

So, interactions beyond the SM%
\footnote{There is, however, a possibility that SM
instanton-induced processes may give a large neutrino-nucleon
cross-section~\cite{Fodor:03}.}
are needed in order to explain possible excess of the UHECR flux.
One possibility is high-energy scattering mediated by
gravitational forces in theories with compact extra
dimensions~\cite{Nussinov:99}-\cite{Emparan:02}. In a number of
papers~\cite{Emparan:02}-\cite{Ahn:03} it was shown that in a
model with extra dimensions the neutrino-nucleon cross section can
be enhanced by a black hole production. The corresponding cross
sections were estimated to be one order of magnitude or more above
the SM predictions \eqref{90} at $E_{\nu} \gtrsim 10^{18}$ eV.

In papers \cite{Alvarez-Muniz:02}, \cite{Kowalski:02} the
opportunities were considered to search for black hole signatures
by using neutrino telescopes such as AMANDA/IceCube, Baikal,
ANTARES or NESTOR. The expected black hole production cross
section is around $10^6$ pb for $M_{BH}^{min} = M_D = 1$ TeV,
where $M_{BH}^{min}$ is a minimal mass of produced black
hole.%
\footnote{The production rate of black holes depends on the number
of extra dimensions and, essentially, on the ratio
$M_{BH}^{min}/M_D$.}

Another possibility, which we will concentrate on, is an
observation of air showers triggered by UHE neutrino interactions.
The technique used for studying extensive air showers of UHECRs or
UHE neutrino is the detection of shower particles by ground
detectors, or the detection of fluorescence light produced by the
shower. The first technique is used by one of the largest
operating AGASA experiment, while the second one was developed for
Fly's Eye (HiRes) detector. The largest project under construction
is the Pierre Auger Observatory~\cite{AUGER}. It will consist of
two sites, each having 1600 particle detectors overlooked by four
fluorescence detectors. For a detail study of extensive air
showers with energy above $10^{18}$ eV, $10$\% of the events will
be detected by both ground array and fluorescence detectors.

It is worth also mentioned space-based experiments EUSO and OWL
which will be sensitive to CRs with energies above $10^{19}$ eV.
The future of the neutrino astronomy may be related with radio
frequency detectors, such as RICE and ANITA.

The neutrino interaction length is given by (in units of km water
equivalent, 1 km we $\equiv 10^5$ g cm$^{-2}$)
\begin{equation}
L_{\nu}(E_{\nu}) \simeq 1.7 \cdot 10^7 \left[
\frac{\text{1pb}}{\sigma_{\nu N} (E_{\nu})} \right] \text{ km we}.
\label{94}
\end{equation}
For typical black hole production cross section $\sigma_{\nu N} =
10^6$ pb, we get $L_{\nu} = 17$ km we. This interaction length is
much larger than the vertical Earth's atmospheric depth, which is
equal to 0.01 km we. The atmospheric depth for neutrinos
transverse (almost) horizontally  is 36 times larger. That is why,
it was proposed to search for uniformly produced quasi-horizontal
showers at ground level~\cite{Berezinsky:75}.%
\footnote{At large zenith angles a background from hadronic cosmic
rays is negligible, since the showers initiated by hadrons are
high in the atmosphere due to very short interaction length of the
proton. Around $10^{20}$ eV, the hadronic mean free path is only
40 g cm$^{-2}$, and gamma-rays of such energy have interactions
lengths of 45-60 g cm$^{-2}$.}

The Fly's Eye and AGASA Collaborations have searched for deeply
penetrating quasi-horizontal air showers, with the depth $L_{sh}
> 2500$ g cm$^{-2}$. The probability of cosmic protons and gamma rays
initiating air showers deeper than $2500$ g cm$^{-2}$ is about
$10^{-9}$. Thus, any shower starting that deep in the atmosphere
is a nice candidate for a neutrino event.

The non-observation of such events puts an upper limit on the
product of the neutrino differential flux, $\Phi_{\nu} = (1/4\pi)
dN_{\nu}/dE_{\nu}$, times neutrino-nucleon cross section. The
Fly's Eye Collaboration gives the bound which can be parametrized
by~\cite{FlyEye}
\begin{equation}
(\Phi_{\nu} \, \sigma_{\nu N})(E_{\nu}) \leq 3.74 \cdot 10^{-42}
\left( \frac{E_{\nu}}{1\text{ GeV}} \right)^{-1.48} \text{
GeV}^{-1} \text{ s}^{-1} \text{ sr}^{-1},
\label{96}
\end{equation}
while the upper limit from Ref.~\cite{Tyler:01} can be recast as
follows
\begin{equation}
(\Phi_{\nu} \, \sigma_{\nu N})(E_{\nu}) \leq 10^{-41} \,
\bar{y}^{-1/2} \left( \frac{E_{\nu}}{1\text{ GeV}} \right)^{-1.5}
\text{ GeV}^{-1} \text{ s}^{-1} \text{ sr}^{-1},
\label{98}
\end{equation}
where $\bar{y}$ is an average fraction of the neutrino's energy
deposited into the shower. The inequalities are valid in the range
$10^8 \text{ GeV} \leq E_{\nu} \leq 10^{11} \text{ GeV}$, {\em
provided $\sigma_{\nu N}(E_{\nu}) \leq 10 \, \mu\text{b}$}.

Let us now estimate the neutrino-nucleon cross section in our
approach. The neutrino scatters off the quarks and gluons
distributed inside the nucleon (see a comment after formula
\eqref{100}). Then the cross section is presented by
\begin{equation}
\sigma^{\nu N}_{in}(s) = \int\limits_{x_{min}}^1 dx \sum_i
f_i(x,\mu^2) \sigma_{in} (\hat{s}),
\label{99}
\end{equation}
where $f_i(x,\mu^2)$ is a distribution of parton $i$ in momentum
fraction $x$, and $\hat{s} = xs$ is an invariant energy of a
partonic subprocess. In our approach, the partonic cross section
$\sigma_{in} (\hat{s})$ is defined via the eikonal \eqref{76}. As
it follows from \eqref{76}, $\chi(\hat{s},b)$ is small at $\hat{s}
\lesssim M_D^2$, and we can put $x_{min} = M_D^2/s$ in \eqref{99}.
At $\hat{s} \geqslant M_D^2$, the main contribution  comes from
the region:
\begin{equation}
b^2  \lesssim b_{max}^2(\sqrt{\hat{s}}) = 4 R_g(\hat{s}) \, [\ln
(\hat{s}/M_D^2) + 1].
\label{100}
\end{equation}
We choose the neutrino energy $E_{\nu}$ to be $10^{17}$ eV,
$10^{18}$ eV, $10^{19}$ eV, $10^{20}$ eV, and  $10^{21}$ eV. The
invariant energy of $\nu N$ collision is then 14 TeV, 43 TeV, 137
TeV, 433 TeV and 1370 TeV, respectively. Since $b_{max}(\sqrt{s} =
1370 \,\text{TeV}) \simeq 3 \cdot 10^{-2}$ GeV$^{-1} = 6 \cdot
10^{-3} \text{ fm}$ (for $2 \leqslant d \leqslant 6$), our
assumption that the neutrino interacts with the proton
constituents is well justified.

We use the set of parton distribution functions (PDFs) from
paper~\cite{Alekhin:02} based on an analysis of existing deep
inelastic data in the next-to-leading order QCD approximation in
the fixed-flavor-number scheme. The extraction of the PDFs is
performed simultaneously with the value of the strong coupling and
high-twist contributions to structure functions. The PDFs are
available in the region $10^{-7} < x < 1$, $2.5 \text{ GeV}^2 <
Q^2 < 5.6 \cdot 10^7 \text{ GeV}^2$~\cite{Alekhin:02}. We take the
mass scale in PDFs to be $\mu = 1/b_{max}(\sqrt{\hat{s}})$, with
$b_{max}$ defined by equation \eqref{100}. The result of our
calculations of $\sigma^{\nu N}_{in}(E_{\nu})$ is presented in
Table~\ref{tab:neutrino}.%
\footnote{The SM contributions to neutrino-nucleon cross sections
are not included in Table~\ref{tab:neutrino}.}
These neutrino-nucleon cross sections do not violate experimental
upper bounds \eqref{96}, \eqref{98}.
\begin{table}[h!t]
\begin{center}
\caption{{\small Inelastic neutrino-nucleon cross section for the
graviton induced scattering at fixed neutrino energy, $E_{\nu}$,
for different numbers of extra dimensions $d$ (in \emph{pbarn}).}}
\bigskip
  \begin{tabular}{||c||c|c|c|c|c||}
  \hline $d$ & 2 & 3 & 4 & 5 & 6
  \\ \hline
  $E_{\nu} = 10^{17} \, \text{eV}$ & 8.63
  $\cdot 10^4$ & 5.63 $\cdot 10^3$ & 5.53 $\cdot 10^2$ & 7.16 $\cdot
  10^1$ & 1.13 $\cdot 10^1$
  \\ \hline
  $E_{\nu} = 10^{18} \, \text{eV}$ & 6.53
  $\cdot 10^5$ & 3.39 $\cdot 10^4$ & 2.47 $\cdot 10^3$ & 2.26 $\cdot
  10^2$ & 2.41 $\cdot 10^1$
  \\ \hline
  $E_{\nu} = 10^{19} \, \text{eV}$ & 4.20
  $\cdot 10^6$ & 2.05 $\cdot 10^5$ & 1.21 $\cdot 10^4$ & 8.59 $\cdot
  10^2$ & 6.99 $\cdot 10^1$
  \\ \hline
  $E_{\nu} = 10^{20} \, \text{eV}$ & 2.05
  $\cdot 10^7$ & 1.32 $\cdot 10^6$ & 7.06 $\cdot 10^4$ & 4.29 $\cdot
  10^3$ & 2.94 $\cdot 10^2$
  \\ \hline
  $E_{\nu} = 10^{21} \, \text{eV}$ & 8.74
  $\cdot 10^7$ & 7.47 $\cdot 10^6$ & 4.56 $\cdot 10^5$ & 2.52 $\cdot
  10^4$ & 1.52 $\cdot 10^3$
  \\ \hline
  \end{tabular}
\label{tab:neutrino}
\end{center}
\end{table}

Note, the total SM cross sections for ($\nu + \bar
\nu$)-scattering defined by formula \eqref{90} are equal to 6.27
$\cdot 10^3$ pb, 1.45 $\cdot 10^4$ pb, 3.35 $\cdot 10^4$ pb, 7.72
$\cdot 10^4$ pb, and 1.78 $\cdot 10^5$ pb, respectively. Thus, the
SM interactions become comparable with (larger than) gravity
interaction for $d = 3 \div 4$ (for $d \geqslant 4 \div 5$),
depending on the neutrino energy $E_{\nu}$.

The number of horizontal hadronic air showers with the energy
$E_{sh}$ larger than a threshold energy $E_{th}$, initiated by the
neutrino-nucleon interactions, is given by
\begin{eqnarray}
&& N_{sh}(E_{sh} \geqslant E_{th}) = T N_A \Big[ \int dE_{\nu} \,
\Phi_{\nu}(E_{\nu}) \, \sigma_{\nu N}^{grav}(E_{\nu}) \, {\cal
A}(E_{\nu}) \, \theta(E_{\nu} - E_{th})  \nonumber \\
&& +  \sum_{i = e, \, \mu, \, \tau} \int dE_{\nu_i} \,
\Phi_{\nu_i}(E_{\nu_i}) \, \sigma_{\nu_i N}^{SM}(E_{\nu_i}) \,
{\cal A}(y_i E_{\nu_i}) \, \theta(y_i E_{\nu_i} - E_{th}) \Big],
\label{102}
\end{eqnarray}
where $N_A = 6.022 \cdot 10^{23} \text{ g}^{-1}$, $T$ is a time
interval, and $\cal A$ is a detector acceptance (in units of
km$^3$ steradian water equivalent). The quantity
$\Phi_{\nu_i}(E_{\nu_i})$ in \eqref{102} is a flux of the neutrino
of type $i$, and $\Phi_{\nu}(E_{\nu}) = \sum_{i = e, \, \mu, \,
\tau} \Phi_{\nu_i}(E_{\nu_i})$.%
\footnote{Both neutrino and antineutrino are everywhere included
in the sum.} The inelasticity $y_i$ defines a fraction of the
neutrino energy deposited into the shower in the corresponding SM
process (see below).

The AGASA acceptance for deeply penetrating quasi-horizontal air
showers with zenith angles $\theta > 75^{\circ}$ can be found in
the second paper of Refs.~\cite{Feng:02}. It rises linearly in
$E_{sh}$ in the interval $10^7 \text{ GeV} < E_{sh} < 10^{10}
\text{ GeV}$, while in the ultrahigh high-energy region the
acceptance is constant and equal to ${\cal A}(E_{sh} \geqslant
10^{10} \text{ GeV}) \approx 1.0 \text{ km}^3 \text{ we} \text{
sr}$~\cite{Feng:02}.

The neutrino acceptance of the Pierre Auger detector is roughly 30
times larger, taking into account the ratio between Auger and
AGASA surface areas. The acceptance of the Auger ground surface
array  has been studied in details in Ref.~\cite{Capelle:98},
while the acceptance of fluorescence detector to neutrino-like air
showers with the large zenith angles was calculated in
Refs.~\cite{Diaz:01}, \cite{Guerard:01}. The Auger observatory
efficiency is high, since the low target density in the atmosphere
is compensated by the very large surface area of the array (each
side of it covers an area of 3000 km$^2$). The highest efficiency
for quasi-horizontal shower detection is expected at $E_{sh} >
10^9 \text{ GeV}$~\cite{Capelle:98}.

The number of extensive quasi-horizontal showers induced by the
neutrinos with energy larger than some threshold energy $E_{th}$,
which can be detected by the array of the southern site of the
Pierre Auger observatory, is presented in
Table~\ref{tab:cosmogenic showers}. The neutrino-nucleon cross
section $\sigma_{\nu N}^{grav}$ in Eq.~\eqref{102} describes the
contributions from the reggeized KK gravitons. The cosmogenic
neutrino flux is from Refs.~\cite{Protheroe:96}, assuming a
maximum energy of $E_{max} = 10^{21}$ eV for the UHECR. The
acceptance of the Auger detector is taken from
Ref.~\cite{Capelle:98} (it is not assumed that shower axis falls
certainly in the array).

\begin{table}[h!t]
\begin{center}
\caption{\small Yearly event rates for nearly horizontal neutrino
induced showers with $\theta_{zenith} > 70^{\circ}$ and $E_{sh}
\geqslant E_{th}$ for cosmogenic neutrino flux from
\cite{Protheroe:96} at three values of threshold energy $E_{th}$.
Number of events corresponds to one side of the Auger ground
array.}
\bigskip
\bigskip
  \begin{tabular}{||c||c|c|c|c|c||}
  \hline
  $d$ & 2 & 3 & 4 & 5 & 6
  \\ \hline
  $E_{th} = 10^8$ \ GeV & 34.88 & 2.00 & 0.32 & 0.21 & 0.20
  \\ \hline
  $E_{th} = 10^9$ \ GeV & 30.21 & 1.66 & 0.21 & 0.12 & 0.12
  \\ \hline
  $E_{th} = 10^{10}$ GeV & 13.16 & 0.74 & 0.062 & 0.025 & 0.022
  \\
  \hline
  \end{tabular}
\label{tab:cosmogenic showers}
\end{center}
\end{table}

The neutrino-nucleon inelastic interactions induced by
gravireggeons remind the SM neutral currents events. We assume
that such events should result in hadronic dominated showers
without leading lepton. That is why, we choose the inelasticity to
be equal to unity%
\footnote{The estimates from Ref.~\cite{Emparan:02} are not
applicable in our case, since in \cite{Emparan:02} an energy loss
in \emph{elastic} neutrino-nucleon cross section induced by
``bare'' gravitons was considered, while we deal with
\emph{inelastic} cross section in the gravireggeon model.}
for the events induced by gravireggeon exchange (the first term in
the RHS of Eq.~\eqref{102}). We have also put $y_e =1$ for the SM
\emph{charged} current interactions initiated by electronic
neutrino,  while for the SM \emph{neutral} interactions initiated
by $\nu_e$ and for $\nu_{\mu}/ \nu_{\tau}$-events we have taken
$y_e = y_{\mu} = y_{\tau} =0.24$, following the calculations
presented in Ref.~\cite{Sigl:98}.

The neutrino event rates are expected to be much higher for the
neutrino fluxes obtained in ``optimistic'' scenarios considered in
Refs.~\cite{Kalashev:02}. As an example, we have presented the
yearly event rates for the Z-burst scenario in Table.~4. One can
see from Table~\ref{tab:zburst showers} that the main contribution
to the shower rate comes from the region of extremely high
neutrino energies ($E_{\nu} > 10^{10}$ GeV). It can be understood
as follows: at UHEs, the neutrino flux times $E_{\nu}$ varies
slowly in $E_{\nu}$ in the Z-burst model (up to $2.5 \cdot
10^{12}$ GeV), while both the acceptance of the Auger array and
``gravitational'' part of the neutrino-nucleon cross section rise
with the neutrino energy (see Table~\ref{tab:neutrino}).

\begin{table}[h!t]
\begin{center}
\caption{\small The same as in Table~\ref{tab:cosmogenic showers},
but for the Z-burst neutrino flux from \cite{Yoshida:98}.}
\bigskip
\bigskip
  \begin{tabular}{||c||c|c|c|c|c||}
  \hline
  $d$ & 2 & 3 & 4 & 5 & 6
  \\ \hline
  $E_{th} = 10^8$ \ GeV & $12.60 \cdot 10^2$ & $11.53 \cdot 10^1$  & 9.26 & 1.90 &
  1.50
  \\ \hline
  $E_{th} = 10^9$ \ GeV & $12.59 \cdot 10^2$ & $11.53 \cdot 10^1$  & 9.26 & 1.90 &
  1.50
  \\ \hline
  $E_{th} = 10^{10}$ GeV & $12.55 \cdot 10^2$ & $11.51 \cdot 10^1$  & 9.20 & 1.85 &
  1.44
  \\
  \hline
  \end{tabular}
\label{tab:zburst showers}
\end{center}
\end{table}

The calculations of the yearly event rates in the energy interval
$10^8 \mbox{ GeV} \leqslant E_{sh} \leqslant 10^{11}$ GeV in the
Z-burst scenario result in 44, 2.7, 0.38, 0.26, and 0.25 for $d =
2, 3, 4, 5$ and 6, respectively. Remembering that combine results
from AGASA and Fly's Eye imply an upper bound of 3.5 at 90\% CL
from quasi-horizontal neutrino events~\cite{Feng:02}, and taking
into account that the AGASA acceptance is roughly 30 times smaller
than the Auger acceptance, we can conclude that the Z-burst
neutrinos do not violate bounds
\eqref{96}, \eqref{98} in our scheme for $d \geqslant 3$.%
\footnote{We do not discuss here \emph{cosmological bounds} on the
number of extra dimensions~\cite{Arkani-Hamed:98} (see also
\cite{Mohapatra:03} and references therein).}

\section{Conclusions}

In the model with compact extra spacial dimensions, we have
calculated the contribution of the KK gravireggeons into the
inelastic cross section of the high energy scattering of both
$D$-dimensional and four-dimensional SM particles. The usually
adopted summing non-reggeized gravitons leads to a divergent sum
in KK-number $n$ (for $D \geqslant 6$). In our approach, on the
contrary, the contribution of gravireggeon with the KK-number $n$
to the eikonal is exponentially suppressed at large $n$. As a
result, the corresponding sum in $n$ is finite, and it can be
analytically calculated.

In the case when the SM fields propagate in all $D$ dimensions,
the dependence of the inelastic cross section on invariant energy
$\sqrt{s}$ appeared to be similar to the upper limit for the total
cross section obtained previously for the SM in the
$D$-dimensional flat space-time without gravity. When, on the
contrary, only gravity lives in extra dimensions, the imaginary
part of the eikonal is derived in a closed form, which depends
(except for $\sqrt{s}$ and the impact parameter $b$) on the number
of extra dimensions $d=D-4$ and their size $R_c$.

We have estimated the event rate for the quasi-horizontal air
showers, induced by the interactions of UHE neutrinos  with
nucleons, which can be yearly detected by the ground array of the
Pierre Auger observatory. It decreases rapidly if $d$ varies from
2 to 5. For $d=4$, we expect 10 events per year for the neutrino
flux predicted in the Z-burst model. For the cosmogenic neutrino
flux, gravireggeon induced interactions do not increase the event
rate significantly with respect to the number of the neutrino
events calculated in the SM, except for the case $d \leqslant 3$,
which is likely to be excluded by the cosmological data.
\vfill\eject

\end{document}